\documentclass[twocolumn,aps,prl]{revtex4}
\usepackage{graphicx}
\begin{document}

\topmargin -5mm
\preprint{This line only printed with preprint option}

\title{Low temperature structural phase transition and incommensurate lattice modulation in the spin gap compound BaCuSi$_2$O$_6$}

\author{E. C. Samulon$^1$, Z. Islam$^2$, S. E. Sebastian$^1$, P. B. Brooks$^3$, M. K.  McCourt Jr.$^3$, J. Ilavsky$^2$, I. R. Fisher$^1$}

\affiliation{$^1$Geballe Laboratory for Advanced Materials and
Department of Applied Physics, Stanford University, Stanford, CA 94305}

\affiliation{$^2$Advanced Photon Source, Argonne National Laboratory,
9700 S. Cass Avenue, Argonne, IL 60439}

\affiliation{$^3$Department of Physics, Stanford University, Stanford, CA 94305}

\begin{abstract}
Results of high resolution x-ray diffraction experiments are presented for single crystals of the 
spin gap compound BaCuSi$_2$O$_6$ in the temperature 
range from 16 to 300 K. The data show clear evidence of a transition from the room temperature
tetragonal phase into an incommensurately modulated orthorhombic structure below $\sim$100 K. 
This lattice modulation is characterized by a resolution limited wave
vector {\bf q}$_{IC}$=(0,$\sim$0.13,0) and its 2$^{nd}$ and 3$^{rd}$ harmonics.
The phase transition is first order and exhibits considerable hysteresis.
This observation implies that the spin Hamiltonian representing the system is more 
complex than originally thought.
\end{abstract}


\date{\today}

\maketitle

\section{\label{sec:level1}Introduction}
BaCuSi$_2$O$_6$ is a quasi-2D compound composed of copper silicate layers separated by Ba ions. 
Within the Cu$_2$Si$_4$O$_{12}$ layers, Cu$^{2+}$ ions are arranged in well-separated, vertical dimers \cite{Finger_1989,Sparta_2004}. 
Consistent with the dimerized structure, 
the material is observed to have a singlet ground state in zero magnetic field, 
with a large gap to the lowest excited triplet states \cite{Sassago_1997}. 
Magnetic fields in excess of $H_{c1}$ $\sim$ 23 T close the spin gap, 
such that cooling in a large applied field results in a state characterized 
by long-range magnetic order \cite{Jaime_2004}. 
At $T$ = 0, a quantum critical point (QCP) at $H_{c1}$ separates the quantum paramagnetic regime from the ordered state. 
Recent experiments probing the critical exponent associated with the approach of the 
phase boundary towards the QCP indicate that 
it is possible to describe the phase transition in terms of Bose-Einstein Condensation (BEC) of delocalized triplets  
down to the lowest measured temperatures of 30 mK \cite{Sebastian_2005a, Sebastian_2005b}. 
Unfortunately, the large spin gap of this material currently precludes 
direct measurement of the magnetic structure in the ordered state. 
For this reason it is particularly important to have a detailed understanding of the low-temperature crystal structure 
in zero magnetic field, since this will have important consequences for the nature of the high-field magnetic 
phase transition. 

To date, the crystal structure of BaCuSi$_2$O$_6$ has only been determined at room temperature and above. 
Sparta and Roth \cite{Sparta_2004} have found evidence for a subtle structural phase transition at 610 K, 
from a high temperature (HT) phase with $I4/mmm$ (no. 139) symmetry, to a room temperature (RT) phase with $I4_{1}/acd$ (no. 142) symmetry. 
However, recent heat capacity and susceptibility measurements indicate the 
presence of an additional first-order phase transition at approximately 100 K \cite{Stern_2005}, 
which is also likely structural in origin. 
In this paper we present results of high resolution x-ray diffraction experiments performed 
at the Advanced Photon Source (APS) on single crystals of BaCuSi$_2$O$_6$ from 16 to 300 K. 
We find clear evidence for a first-order structural phase transition at approximately 100 K to a low temperature (LT) crystal structure, 
characterized by an orthorhombic distortion of the RT tetragonal structure, 
and the appearance of an additional incommensurate lattice modulation. 
We discuss the origin of this effect and the consequences for the high field ordered state.

\section{Experimental methods}
Single crystal samples of BaCuSi$_2$O$_6$ were grown via a slow-cooling flux technique, 
as described elsewhere \cite{Sebastian_2005a}. 
The crystals were well-formed, and had a small mosaic spread of $\sim0.02^\circ$. 
High resolution x-ray diffraction experiments were performed on the X-ray
Operations and Research 1BM bending magnet beamline \cite{Lang_1999} at the 
APS, Argonne National Laboratory. 
This study was done with a focused beam of 20.016 keV x-rays.
A Si(111) analyzer crystal was used in order to suppress background and
improve resolution. The sample was cooled using a closed-cycle refrigerator,
with a base temperature of $\sim$16K. A Si-diode sensor was placed about 1 cm below
the sample to measure its temperature. 

Susceptibility measurements were made using a commercial Quantum Design MPMS5 SQUID magnetometer, 
for a field of 5000 Oe aligned parallel and perpendicular to the crystalline $c$-axis. 
Data were taken for both increasing and decreasing temperatures. 
Two different instruments were used, with slight differences in the cooling rate 
and degree of undercooling on approaching a temperature set point.

\section{Results}
The temperature dependence of the susceptibility of BaCuSi$_2$O$_6$ follows a form typical 
for weakly coupled spin dimer systems, 
and can be fit by a dimer model with a spin gap of 4.45 meV, 
as has been previously reported by Sassago and co-workers \cite{Sassago_1997}. 
Superimposed on this behavior, a weak jump in the susceptibility can be observed 
in the temperature range between 85 and 102 K (Figure 1). 
The temperature at which the jump occurs depends on the thermal history of the sample, 
indicative of a first order phase transition in this temperature range. 

\begin{figure}
\includegraphics[width=8.5cm]{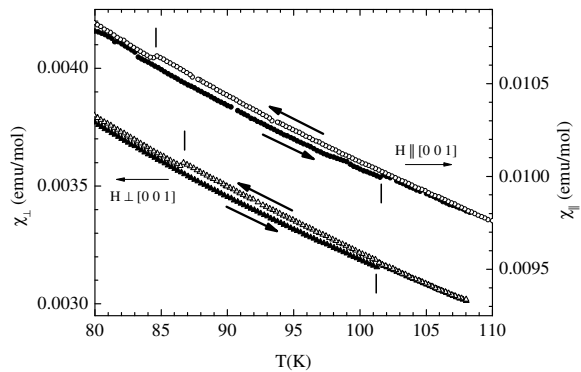}
\caption{Temperature dependence of the susceptibility of BaCuSi$_2$O$_6$ between 80 and 110 K, 
for an applied field of 5000 Oe oriented parallel and perpendicular to the crystalline $c$ axis (circles and triangles respectively). 
Data are shown for warming (solid symbols) and cooling cycles (open symbols), indicated by arrows. 
Jumps in susceptibility associated with the first order phase transition are indicated by vertical lines.
Differences in the cooling rate result in the 
difference in the lower temperature jump for the two data sets.}
\label{Fig1}
\end{figure}
At room temperature, well above the transition observed in the susceptibility, 
charge Bragg peaks were observed consistent with the tetragonal structure previously reported by Sparta \cite{Sparta_2004}. 
However, at 16 K the even-order ($h$,0,0) Bragg peaks were observed to have split along the
longitudinal direction as shown in Fig. 2. In addition, incommensurate (IC)
satellite peaks appeared on either side of the split Bragg peaks. A careful set
of measurements revealed that the splitting is consistent with transformation
twins of an orthorhombic (or weakly monoclinic) lattice. These twins are rotated with respect to
each other about the 2-fold $c$-axis by $\sim89^\circ$ as indicated in the inset to 
Fig. 2. From the splitting we found the degree of orthorhombicity defined as
$\Delta_o=\frac{|a_o-b_o|}{\frac12 (a_o+b_o)}\approx0.2\%$, where
$a_o$ and $b_o$ are the orthorhombic lattice parameters ($a_0 < b_0$; Fig. 3).
The $c$-axis lattice parameter also exhibits a weak reduction ($\sim$0.5\%) below the transition temperature. 
The incommensurate peaks are characterized by a wavevector
{\bf q}$_{IC}=(0,\sim0.13,0)$ and its $2^{nd}$ and $3^{rd}$ harmonics.
These IC peaks are
resolution limited in all three directions implying a fully
3-dimensionally ordered modulated structure for the LT phase. 
The intensity of the fundamental satellite is some $\sim10^5$ times
weaker than that of the (0,8,0) Bragg peak. 
\begin{figure}
\includegraphics[width=8.5cm]{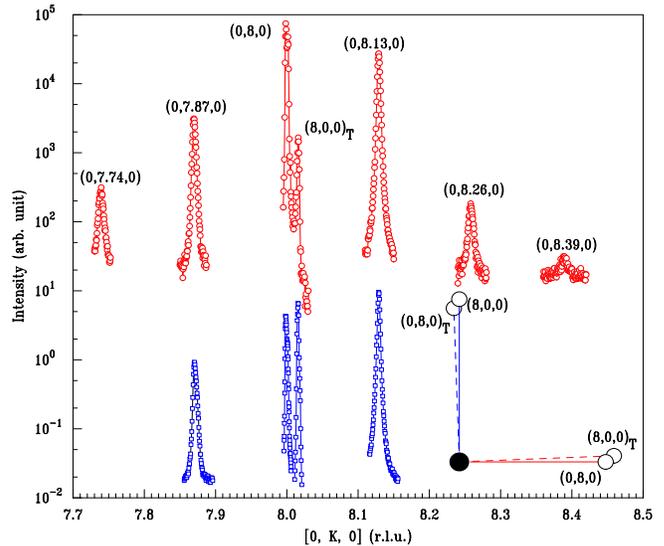}
\caption{(Color on line). Reciprocal lattice scans (in red) along the {\bf b*} axis showing the
orthorhombic splitting of (8,0,0) 
Bragg peak at 16 K. IC diffraction peaks and
their higher harmonics are observed on either side of the Bragg point. Note that
the corresponding peaks on two sides are equidistant from (0,8,0), not
from (8,0,0)$_T$ of the twin partner. Schematic diagram in the lower right corner 
illustrates split peaks from the twins. 
Scans along the orthogonal direction also reveal similar satellites (in blue; data were displaced down for clarity). 
In this case, the satellite peaks are associated with the (0,8,0)$_T$ twin. 
Central Bragg peaks have been scaled down by a factor of $\sim$10$^5$
to display on the same scale.
}
\label{Fig2}
\end{figure}

The orthorhombic splitting and the intensity of the IC (0,8.13,0) peak were
measured as a function of increasing temperature (Fig. 4). Both the splitting
and the intensity of IC peak remain nearly constant up to 103.8 K and disappear
above that temperature.  Within the resolution of our measurements we did not
observe any change of {\bf q}$_{IC}$ up to 103.8 K. 
An extensive search at 110 K confirmed the absence of both the orthorhombic splitting
and the IC modulation. 
On cooling, both the splitting and the IC peaks reappear below $\sim91 K$
indicating hysteretic effects, similar to what is observed in the susceptibility (Fig. 1). 
The hysteresis of the transition is spectacularly
demonstrated by the behavior of the IC peak with thermal cycling through the
transition (inset, Fig. 4). On warming from 16 K to 103.1 K, just below
the transition, the peak is clearly observed (red circles). By raising the sample T to 104.6 K,
just above the transition, and subsequently cooling down to 100 K we were
not able to recover the IC peak at 102.8 K (black squares). However, cooling down to 80 K and subsequently heating 
was sufficient to observe the peak up to 103 K on warming again (blue triangles). From such measurements
we deduced that the transiton is of first order, with transition temperatures
($T_c$) of $104.1\pm0.2 K$ and $91K\pm1 K$ on warming and cooling, respectively.
The observed transition width is remarkably sharp ($ \frac{\Delta T}{T_c}\sim0.003$),
which is a measure of high quality and purity of the bulk single crystal.
\section{Discussion}
Low-T incommensurately modulated structures are not uncommon in silicates. 
Such modulations are usually due to a displacive phase 
transition of the lattice involving rotations of SiO$_4$ tetrahedra without 
substantial internal distortions of the tetrahedra themselves \cite{Hoeche_2002, Giddy_1993, Hammonds_1996}. 
BaCuSi$_2$O$_6$ can be thought of as a `vierer' single ring cyclosilicate \cite{Liebau_1985}, 
in which  isolated Si$_4$O$_{12}$ rings (composed of 4 corner sharing SiO$_4$ tetrahedra) 
are linked by CuO$_4$ groups in the $ab$-plane, 
and separated by Ba cations between successive layers. 
It is likely that SiO$_4$ tetrahedra 
in this compound rotate and twist as a unit with respect to each other
causing a lattice modulation involving both Ba and/or Cu atoms. 
The harmonic content (Figure 2) implies a subtle squaring-up of the modulated structure, 
but a detailed description of the atomic motions awaits a complete determination of the LT structure, which is 
beyond the scope of this initial survey. 
 
\begin{figure}
\includegraphics[width=8.5cm]{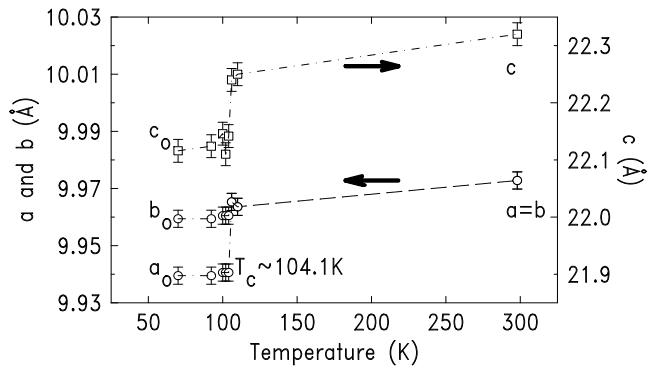}
\caption{Temperature dependence of the lattice parameters for a warming cycle.}
\label{Fig3}
\end{figure}
The susceptibility data shown in Figure 1 demonstrate that the change in crystal structure between the 
RT and LT phases of BaCuSi$_2$O$_6$ affects the magnetism of the system, albeit very weakly. 
These changes are presumably associated with subtle changes in the 
superexchange parameters coupling the spins in the lattice. Without a detailed structural model, it is difficult
to predict how the intradimer (J) and interdimer (J$'$) exchange constants are affected, but we
can anticipate that both will be modulated to some degree. 
We can therefore be fairly sure that the full spin Hamiltonian describing the system is slightly more complex than initially thought. 
We should note, however, that this structural study cannot provide an energy scale for the variations 
in the exchange partameters beyond suggesting that they are  rather small, 
given that the intensity of the incommensurate satellite peaks is several orders of magnitude lower than 
the Bragg peaks associated with the average crystal structure. 
\begin{figure}
\includegraphics[width=8.5cm]{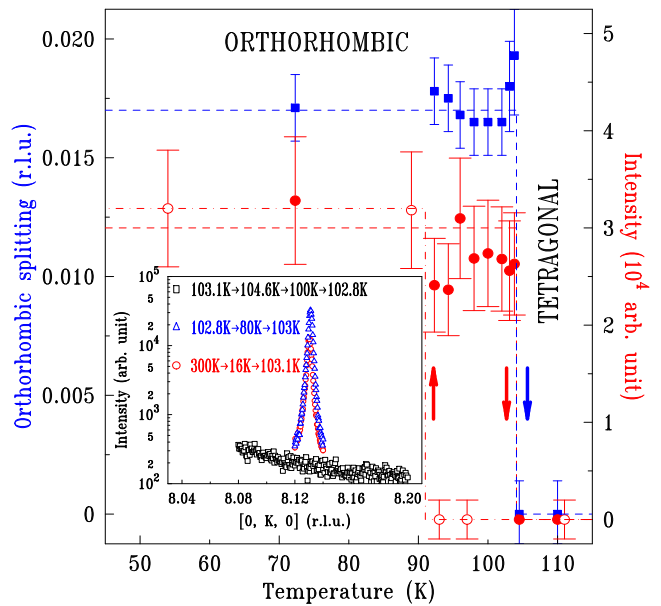}
\caption{(Color on line.) Temperature dependence of the orthorhombic splitting of
the lattice (blue squares, left axis) and the intensity of fundamental IC modulation peak
observed at (0,8.13,0) (red closed circles, right axis) measured on warming. The intensity of the
IC satellite on cooling is shown as open red circles. Note the hysteresis. Inset displays scans
through the satellite region for three different thermal cycling profiles through the
transition, discussed in the main text. The data in the inset correspond to the final temperature
of the given sequence of temperature set points.}
\label{Fig4}
\end{figure}

The change in symmetry at the structural phase transition can also affect the magnetism via spin-orbit coupling. 
The HT ($I4/mmm$) structure has a center of inversion symmetry between the two Cu ions that comprise a dimer unit. 
However, the RT ($I4_{1}/acd$) and (presumably) LT structures do not, 
so additional terms $\vec D\cdot \vec S_1\times \vec S_{2}$ (where $\vec S_1$ and $\vec S_2$ label spins in one dimer unit) 
due to antisymmetric Dzyaloshinski-Moriya (DM) exchange 
are not forbidden in the spin Hamiltonian for these phases. 
This is the lowest order effect by which spin-orbit coupling can introduce terms 
to the spin Hamiltonian that break axial U(1) symmetry (an essential prerequisite for BEC \cite{Sirker_2004}), 
since anisotropy in the $g$-tensor simply rescales the critical field $H_{c1} = \Delta/g\mu_B$. 
In the RT structure, copper dimers reside on axes of improper rotation $\bar4$ (an inversion tetrad), 
which within a single Cu$_2$Si$_4$O$_{12}$ sheet closely resembles an axis of 4-fold symmetry. 
According to the standard symmetry rules for DM exchange \cite{Yosida_book}, 
the corresponding $\vec D$-vector for the RT phase would then point along the 4-fold axis.
Hence, for the RT structure it is anticipated that the U(1) symmetry of the spin Hamiltonian will be preserved 
for magnetic fields aligned parallel to the crystalline $c$-axis. 
Without a complete structural model for the LT phase we have less insight to the symmetry of the dimers. 
However, we note that the structural distortion is slight, 
which implies that there will be minimal changes in the $\vec D$ vector. 

In principle, the orthorhombicity of the LT structure can introduce an additional anisotropy 
to the spin Hamiltonian via a second order effect associated with spin-orbit coupling. 
For a tetragonal structure, the {\it x} and {\it y} components of the interdimer exchange 
coupling J$'$ are perforce equal, but for an orthorhombic structure this does not have to
be the case. The resulting symmetric anisotropy is quadratic in the spin-orbit interaction, 
({\it i.e.} only appears in second order corrections) and is therefore even smaller than
any DM terms.

While the symmetry of the crystal lattice implies that the spin-orbit effects discribed above are possible, 
the magnitude of those terms must be measured by separate techniques.
Ongoing ESR experiments indicate that the energy scale associated with the DM terms in both the LT and RT phases is negligible \cite{Hill_2005}.
Furthermore, the observed critical scaling exponents associated with the phase boundary imply that any U(1) symmetry 
breaking terms that do enter the Hamiltonian must be on a significantly lower energy scale than the temperatures 
over which the magnetic ordering transition has been observed.

Finally, we note that the observation of an incommensurate structural modulation helps to understand 
results of preliminary high-field NMR experiments on this material. 
In these experiments, Stern and coworkers have studied the NMR spectrum of $^{29}$Si nuclei 
at 0.04 K for fields just below $H_{c1}$ (23.4 T), and just above $H_{c1}$ (24.4 T)\cite{Horvatic_2005}. 
Rather surprisingly, they find that the sharp NMR line below $H_{c1}$ is substantially broadened for fields above $H_{c1}$. 
Within a simple interpretation of the proposed triplet BEC, one would instead anticipate a splitting of the 
NMR lines corresponding to a doubling of the unit cell associated with the expected  $(\pi/a,\pi/a)$ wave vector of the 
staggered magnetization\cite{Horvatic_2005, Sebastian_2005b}. 
The incommensurate structural modulation provides means to understand this observation 
without necessarily having to introduce a more complex magnetic structure. 
At a temperature of 0.04 K for fields below $H_{c1}$ there are essentially no thermally populated triplets, 
the uniform magnetization is zero, and the Si nuclei all see the same local magnetic field resulting in a relatively narrow NMR line. 
However, for fields above $H_{c1}$, there is a finite magnetization which causes a substantial Knight shift. 
Due to the incommensurate structure, each Si nucleus would see a slightly different magnetic field, 
broadening the NMR lineshape. That is to say that the broadened NMR lines seen for fields above $H_{c1}$ 
do not necessarily correspond to an incomensurate {\it magnetic} structure, 
but more likely reflect the underlying incommensurate {\it crystal} structure of the LT phase. 
Further NMR experiments are called for to experimentally distinguish these two effects.  

\section{Conclusions}
In summary, our experiments have unambiguously shown the presence of a low-temperature structural phase transition in 
BaCuSi$_2$O$_6$, corresponding to an orthorhombic distortion of the RT $I4_{1}/acd$ structure and 
accompanied by an incommensurate modulation. 
The transition is first order, having a substantial hysteresis 
spanning approximately 80 to 105 K, depending on details of the warming or cooling cycles. 
The lower symmetry and incommensurate modulation of the LT phase imply that the spin Hamiltonian describing the magnetic properties of the system is more complex than previously thought.
ESR and inelastic neutron scattering experiments are currently in progress to determine the energy scale for additional relevant terms. 
Finally, the observation of an incommensurate structural modulation at low temperatures provides important insight to recent high field NMR measurements for this compound. 
\section{Acknowledgments}
The authors are grateful to C. Batista, N. Harrison, M. Jaime, S. Hill. K. Sparta, R. Stern
and P. L. Lee for useful discussions.
The use of the APS is supported by the U.S. DOE, Office
of Science, under Contract No. W-31-109-ENG-38. This work is supported by
the National Science Foundation, DMR-0134613. 
I. R. F. acknowledges support from the Alfred P. Sloan Foundation 
and S. E. S. from the Mustard Seed Foundation.

\end{document}